\documentclass[aps,showpacs,preprint]{revtex4}
\newcommand{\zz}[1]{{}}

\begin{document}

\title{Number-phase Wigner function on extended Fock space}

\author{Kiyotaka Kakazu}

\affiliation{Department of Physics and Earth Sciences 
University of the Ryukyus, Okinawa 903-0213, Japan}

\begin{abstract}
On the basis of the phase states, we present the correct integral 
expressions of the two number-phase Wigner functions discovered 
so far.  These correct forms are derived from those defined in 
the extended Fock space with negative number states.  The 
analogous conditions to Wigner's original ones cannot lead to the 
number-phase function uniquely.  To show this fact explicitly, 
we propose another function satisfying all these conditions.  
It is also shown that the ununiqueness of the number-phase Wigner 
function result from the phase-periodicity problem.  
\end{abstract}

\pacs{03.65.Ca, 42.50.-p}


\maketitle

\section{Introduction}

The Wigner function, first introduced by Wigner, gives the phase-space 
formulation of quantum mechanics for a pair of position and momentum 
that have continuous spectra \cite{wign3249,ocon8145,bala8447,hill8421}.  
Since there is a correspondence between quantum observables and 
their classical-like functions in this formalism, physical quantities 
such as the expectation value of any observable can be calculated 
using the quasiprobability distribution (Wigner function).  

Requiring the analogous properties of Wigner's original function 
\cite{hill8421} and introducing some additional assumptions, Vaccaro 
\cite{vacc9574} has defined a number-phase Wigner function 
$S_1(n,\theta) = \textrm{Tr}[\hat S_1(n,\theta)\hat\rho]$ for a density 
matrix $\hat\rho$, where the Wigner operator $\hat S_1(n,\theta)$ has 
the form \zz{a1}
\begin{equation}
  \hat S_1(n,\theta)
    = \frac{1}{\pi}\int_0^{2\pi}\!\!\!d\xi\, e^{2in\xi}
      \left(\frac{1+e^{-i\xi}}{2}\right)
      |\theta-\xi;p\rangle\langle\theta+\xi;p|.\label{a1}
\end{equation}  
Here $|\theta;p\rangle$ is the phase state \zz{a2}
\begin{equation}
  |\theta;p\rangle 
   = \frac{1}{\sqrt{2\pi}}\sum_{n=0}^\infty e^{in\theta}|n\rangle, 
                                               \label{a2}
\end{equation}
where the symbol \lq\lq p" stands for a physical state. 
The function $S_1$ gives a representation of the state $\hat\rho$ that 
displays the underlying photon-number and phase properties.  

Several authors \cite{berr7737,muku7982,biza9455,ruzz0233} have studied 
the rotational Wigner function for rotation angle and angular momentum.  
In particular, Bizarro \cite{biza9455} derived the function starting 
from six \lq\lq natural" conditions, which constitute an appropriate 
way for constructing the function.  
If we interpret rotation angle and angular momentum as (extended) 
phase and (extended) number operators, respectively, the rotational 
Wigner function becomes a number-phase distribution function in the 
extended space.  Moreover, restricting the resulting function to the 
physical space, we obtain another number-phase Wigner operator: \zz{a3}
\begin{equation}
 \hat S_2(n,\theta) 
   = \frac{1}{\pi}\int_{-\pi/2}^{\pi/2}\!\!\!d\xi\, e^{2in\xi}
     |\theta-\xi;p\rangle\langle\theta+\xi;p|, \label{a3}
\end{equation}

Introducing a superoperator (acting on any operator) and some relations 
for the phase-number pair corresponding to the position-momentum pair, 
Ban \cite{ban9569} has obtained an alternative number-phase distribution 
function, which is different from $S_1$ and $S_2$.  
We do not consider Ban's function here, because this function 
is defined without the above analogous conditions and does not satisfy 
even phase-shift condition, one of basic requirements.  

At present we have at least three different number-phase functions.  
All these functions are calculated by using the number states 
to avoid $2\pi-$periodicity problem of phase variable.  As a 
consequence, as is shown in Eqs.\ (\ref{a1}) and (\ref{a3}), the 
phase-periodicity structure of the functions is not clear.  The 
relationship between these functions is not clear either.  Moreover, 
the integral representations (\ref{a1}) and (\ref{a3}) have a problem 
called an endpoint problem \cite{kazu0255}: that is, we have \zz{a3a}
\begin{eqnarray}
 &&\hat S_1(n,\theta)|\theta;p\rangle
    = \frac{1}{2\pi}\int_0^{2\pi}\!\!\!d\xi\, e^{2in\xi}
      \left(\frac{1+e^{-i\xi}}{2}\right)
      \Big[\delta(\xi)+\delta(\xi-2\pi)\Big]
      |\theta-\xi;p\rangle + \cdots,         \nonumber\\
 &&\hat S_2(n,\theta)|\theta-\pi/2;p\rangle
    = \frac{1}{2\pi}\int_{-\pi/2}^{\pi/2}\!\!\!d\xi\, e^{2in\xi}
      \delta(\xi+\pi/2)|\theta-\xi;p\rangle + \cdots, \label{a3a}
\end{eqnarray}
which are ambiguous.  In deriving Eq.\ (\ref{a3a}), we have used \zz{a3b}
\begin{equation}
  \langle\theta;p|\theta';p\rangle 
   = \frac{1}{2}\sum_{n=-\infty}^\infty\delta(\theta-\theta'-2n\pi)
     + \frac{i}{4\pi}\cot\left(\frac{\theta-\theta'}{2}\right)
     + \frac{1}{4\pi}.      \label{a3b}
\end{equation}

This problem is due to the fact that it is not always possible to 
integrate a generalized function over a finite region.  An integral 
over a finite region should be converted into an integral over 
$(-\infty, \infty)$.  For this purpose, we must use a generalized 
periodic function and a unit function \cite{kazu0255,ligh58}.  Thanks 
to the unit function, the phase and number variables can be treated 
freely as the position and momentum variables without bothering with the 
phase-periodicity problem.  

The purpose of the present article is to present the correct 
representations for the operators $\hat S_1$ and $\hat S_2$, and to 
find the relationship between them by using the six \lq\lq natural" 
conditions.  The correct integral expressions can be obtained by 
making use of the phase states.  Then, it becomes clear that 
$\hat S_1$ and $\hat S_2$ cannot be defined uniquely, although 
Bizarro  \cite{biza9455} states that $\hat S_2$ is determined uniquely 
by six \lq\lq natural" conditions.  In fact, there are infinite 
possibilities of defining number-phase Wigner operators satisfying 
these conditions.  

In the next section, the six conditions for determining the Wigner 
operator are considered, using the phase states in the extended Fock 
space.  Then, the nonuniqueness of the Wigner operator becomes clear.  
In section 3, we show that the basic properties which the Wigner 
operator should satisfy can be derived from a part of six conditions.  
The correct forms of the operators $\hat S_1$ and $\hat S_2$ are 
given in section 4.  We end with a conclusion in section 5.

\section{Conditions for the Wigner operator}

In this section, we first give a brief review of the quantum phase in 
the extended Fock space to overcome the endpoint problem.  The 
six \lq\lq natural" conditions and the nonuniqueness of the Wigner 
operator are considered, using mainly the phase states in the extended 
space.  

\subsection{Phase states}

The basis of the extended Fock space  is given by 
$\{|n\rangle\}$ $(n=0, \pm1, \pm2, \cdots)$ \cite{kazu0255}, where the 
(extended) phase states and the number states are defined by \zz{a4}
\begin{equation}
  |\theta\rangle 
   = \frac{1}{\sqrt{2\pi}}\sum_{n=-\infty}^\infty\,
     e^{in\theta}|n\rangle, \quad
|n\rangle
   = \frac{1}{\sqrt{2\pi}}\int_{-\infty}^{\infty}\!\!\!du(\theta)\, 
     e^{-in\theta}|\theta\rangle,       \label{a4}
\end{equation}
where $du(\theta)= U(\theta)d\theta$.  Here $U(\theta)$ is called a 
rapidly decreasing unit function \cite{ligh58} satisfying \zz{a5}
\begin{equation}
  U(\theta) = 0\quad (|\theta|\ge 2\pi), \quad
  \sum_{n=-\infty}^\infty U(\theta+2n\pi) = 1, \quad
  \sum_{n=-\infty}^\infty U^{(k)}(\theta+2n\pi) = 0, \label{a5}
\end{equation}
where $ U^{(k)}$ is the $k$th derivative of $U$.  An example of a unit 
function could be given by \cite{ligh58} \zz{a5a}
\begin{equation}
  U(\theta) = \frac{1}{K}\int_{|\theta|}^{2\pi}\!\!\!dt\, 
              \exp\left[\frac{-4\pi^2}{t(2\pi-t)}\right], \quad
  (|\theta|\le 2\pi)\label{a5a}
\end{equation}
where \zz{a5b}
\begin{equation}
  K = \int_0^{2\pi}\!\!\!dt\, 
      \exp\left[\frac{-4\pi^2}{t(2\pi-t)}\right]. \label{a5b}
\end{equation}
The states $|\theta\rangle$ and $|n\rangle$ satisfy the 
orthonormality relations \zz{a6}
\begin{equation}
  \langle n|m\rangle = \delta_{nm}, \quad
  \langle\theta|\theta'\rangle
    = \sum_{n=-\infty}^\infty\delta(\theta-\theta'+2n\pi)
    \equiv\delta_{2\pi}(\theta-\theta'). \label{a6}
\end{equation}
Here $\delta_{2\pi}(\theta-\theta')$ is a periodic generalized function 
with a period of $2\pi$.  
The phase operator $\hat\theta$ and the number operator $\hat N$ can be 
represented as \cite{kazu0255} \zz{a7}
\begin{equation}
 \hat\theta
   = \int_{-\infty}^{\infty}\!\!\!du(\theta)\, [\theta]
     |\theta\rangle\langle\theta|, \quad
 \hat N
   = \sum_{n=-\infty}^\infty n |n\rangle\langle n|, \label{a7}
\end{equation}
where $[\theta]$ is a sawtooth wave, a periodic function with a 
period of $2\pi$ of the phase variable $\theta$: 
$[\theta] = \theta\ (0\le\theta<2\pi)$ [see \cite{kazu0255}].  

Let us next consider the number-phase Wigner function $W(n,\theta) = 
\textrm{Tr}[\hat W(n,\theta)\hat\rho]$ in the extended Fock space, 
where the number-phase Wigner 
operator can be expressed in terms of the number states and the phase 
states: \zz{a8}
\begin{eqnarray}
  \hat W(n,\theta)
   &=& \sum_{k=-\infty}^\infty\sum_{\ell=-\infty}^\infty
     |k\rangle\langle k|\hat W(n,\theta)|\ell\rangle\langle\ell|
                                        \nonumber\\
   &=& \int_{-\infty}^\infty\!\!\!du(\xi,\xi')\, 
     |\xi\rangle\langle\xi|\hat W(n,\theta)|\xi'\rangle\langle\xi'|,
                                           \label{a8}
\end{eqnarray}
where $du(\xi,\xi')=du(\xi)du(\xi')$.  
In Eq.\ (\ref{a8}), the completeness relations \zz{a9}
\begin{equation}
  \sum_{n=-\infty}^\infty|n\rangle\langle n| = 1,\qquad
  \int_{-\infty}^\infty\!\!\!du(\theta)\, 
  |\theta\rangle\langle\theta| = 1         \label{a9}
\end{equation}
has been used.  

For later convenience, we give here some useful formulas: \zz{a7a1-a7a4}
\begin{eqnarray}
 &&\int_{-\infty}^{\infty}\!\!\!du(\xi)\, \delta_{2\pi}(\xi-\theta)
   \langle\xi|\psi\rangle
     = \langle\theta|\psi\rangle, \label{a7a1}\\
 &&\int_{-\infty}^{\infty}\!\!\!du(\xi)\, \delta'_{2\pi}(\xi-\theta)
   \langle\xi|\psi\rangle
     = - \frac{\partial}{\partial\theta}\langle\theta|\psi\rangle,
                               \label{a7a2}\\
 &&\int_{-\infty}^\infty\!\!\!du(\xi)\, |\xi+\Delta\rangle
   \langle\xi+\Delta|
    = \int_{-\infty}^\infty\!\!\!du(\xi)\, |\xi\rangle\langle\xi|,
                               \label{a7a3}\\
 &&\int_{-\infty}^\infty\!\!\!d\xi\, U(\theta-\xi)\langle\xi|\psi\rangle
   = \int_{-\infty}^\infty\!\!\!du(\xi)\, \langle\xi|\psi\rangle,
                               \label{a7a4}
\end{eqnarray}
where $|\psi\rangle$ is any state, $|\xi\rangle$ any phase state 
and $\Delta$ a real constant.  Proofs of the formulas (\ref{a7a3}) and 
(\ref{a7a4}) are given in the appendix.  Using Eq.\ (\ref{a7a1}), 
we easily obtain $\hat\theta|\theta\rangle = [\theta]|\theta\rangle$; 
that is, $[\theta]$ $(0\le[\theta]<2\pi)$ is an eigenvalue of the 
phase operator $\hat\theta$.  Note that we can treat the operators 
$\hat\theta$ and $\hat N$ quite easily in the extended space, as 
as seen in equations (\ref{a3b}) and (\ref{a6}).  

\subsection{Six conditions}

Following Bizarro \cite{biza9455} and Vaccaro \cite{vacc9574}, we 
introduce the six conditions to determine the Wigner operator 
$\hat W$.  

(i) The operator $\hat W(n,\theta)$ should be Hermitian and so 
$W(n,\theta)$ should be real.  Thus we have \zz{a10}
\begin{equation}
  \langle k|\hat W(n,\theta)|\ell\rangle
    = \langle\ell|\hat W(n,\theta)|k\rangle^*,\quad
  \langle\xi|\hat W(n,\theta)|\xi'\rangle
    = \langle\xi'|\hat W(n,\theta)|\xi\rangle^*.    \label{a10}
\end{equation}

(ii) $W(n,\theta)$ gives the probability distributions for $n$ and 
$\theta$ \zz{a11,a12}
\begin{eqnarray}
  &&\int_{-\infty}^\infty\!\!\!\!du(\theta)\, W(n,\theta)
     = \langle\psi|n\rangle\langle n|\psi\rangle, 
                                              \label{a11}\\
  &&\sum_{n=-\infty}^\infty W(n,\theta)
     = \langle\psi|\theta\rangle\langle\theta|\psi\rangle.
                                              \label{a12}
\end{eqnarray}
We thus require that \zz{a13,a14}
\begin{eqnarray}
  &&\int_{-\infty}^\infty\!\!\!\!du(\theta)\,  \hat W(n,\theta)
     = |n\rangle\langle n|,                   \label{a13}\\
  &&\sum_{-\infty}^\infty \hat W(n,\theta)
     = |\theta\rangle\langle\theta|.          \label{a14}
\end{eqnarray}

(iii) $W(n,\theta)$ should be Galilei invariant with respect to 
displacements in phase $\theta$ and number $n$.  For phase shift 
of $\Delta$ we have \zz{a15}
\begin{equation}
  \langle\xi|\hat W(n,\theta)|\xi'\rangle
   = \langle\xi+\Delta|\hat W(n,\theta+\Delta)|\xi'+\Delta\rangle,
                                              \label{a15}
\end{equation}
which leads to \zz{a16}
\begin{eqnarray}
  \langle k|\hat W(n,\theta)|\ell\rangle
  &=&\int_{-\infty}^\infty\!\!\!du(\xi,\xi')\, 
     \langle k|\xi\rangle\langle\xi|\hat W(n,\theta)|\xi'\rangle
     \langle\xi'|\ell\rangle   \nonumber\\
  &=&\int_{-\infty}^\infty\!\!\!du(\xi,\xi')\, 
     e^{-i(k-\ell)\Delta}\langle k|\xi+\Delta\rangle
     \langle\xi|\hat W(n,\theta)|\xi'\rangle
     \langle\xi'+\Delta|\ell\rangle  \nonumber\\
  &=& e^{-i(k-\ell)\Delta}\langle k|\hat W(n,\theta+\Delta)|\ell\rangle.
                                              \label{a16}
\end{eqnarray}
Setting $\Delta=-\theta$, we have \zz{a17}
\begin{equation}
  \langle k|\hat W(n,\theta)|\ell\rangle
   = e^{i(k-\ell)\theta}\langle k|\hat W(n,0)|\ell\rangle.
                                              \label{a17}
\end{equation}
In Eq.~(\ref{a16}) we have used the equality for the completeness 
relation (\ref{a7a3}).  

From invariance with respect to the number shift $m$, it follows 
that \zz{a20}
\begin{equation}
  \langle k|\hat W(n,\theta)|\ell\rangle
   = \langle k+m|\hat W(n+m,\theta)|\ell+m\rangle.
                                              \label{a20}
\end{equation}
Hence we have \zz{a21}
\begin{equation}
  \langle k|\hat W(n,\theta)|\ell\rangle
   = \langle k-n|\hat W(0,\theta)|\ell-n\rangle
                                              \label{a21}
\end{equation}
and \zz{a22}
\begin{equation}
  \langle\xi|\hat W(n,\theta)|\xi'\rangle
   = e^{-in(\xi-\xi')}\langle\xi|\hat W(0,\theta)|\xi'\rangle.
                                              \label{a22}
\end{equation}

(iv) The transition probability between the states $|\psi\rangle$ and 
$|\psi'\rangle$ should be given, in terms of the respective 
$W(n,\theta)$ and $W'(n,\theta)$, by \zz{a23}
\begin{equation}
 2\pi\sum_{n=-\infty}^\infty\int_{-\infty}^\infty\!\!\!du(\theta)
 \, W(n,\theta)W'(n,\theta)
   = |\langle\psi|\psi'\rangle|^2.            \label{a23}
\end{equation}

(v) $W(n,\theta)$ should be invariant with respect to number-phase 
reflection.  From this condition we have \zz{c17}
\begin{equation}
  \langle\xi|\hat W(n,\theta)|\xi'\rangle
   = \langle-\xi|\hat W(-n,-\theta)|-\xi'\rangle, \label{c17}
\end{equation}
which is equivalent to \zz{c18}
\begin{equation}
  \langle k|\hat W(n,\theta)|\ell\rangle
   = \langle-k|\hat W(-n,-\theta)|-\ell\rangle.   \label{c18}
\end{equation}

(vi) $W(n,\theta)$ should be invariant with respect to time reversal, 
i.e., $\langle k|\psi\rangle \to \langle k|\psi\rangle^*$ and 
$\langle k|\hat W(n,\theta)|\ell\rangle \to 
\langle-k|\hat W(-n,\theta)|-\ell\rangle$.  We then have \zz{c20}
\begin{eqnarray}
  W(n,\theta)
   &=& \sum_{k,\ell}\langle\psi|k\rangle^*
     \langle-k|\hat W(-n,\theta)|-\ell\rangle\langle\ell|\psi\rangle^*
                                              \nonumber\\
   &=& \sum_{k,\ell}\langle\psi|k\rangle
     \langle-\ell|\hat W(-n,\theta)|-k\rangle\langle\ell|\psi\rangle
                                              \label{c20}
\end{eqnarray}
for any state $|\psi\rangle$.  Hence \zz{c21}
\begin{equation}
  \langle k|\hat W(n,\theta)|\ell\rangle
   = \langle-\ell|\hat W(-n,\theta)|-k\rangle,
                                              \label{c21}
\end{equation}
which leads to \zz{c22}
\begin{equation}
  \langle\xi|\hat W(n,\theta)|\xi'\rangle
   = \langle\xi'|\hat W(-n,\theta)|\xi\rangle.
                                              \label{c22}
\end{equation}

\subsection{Nonuniqueness of the Wigner operator}

We next apply the above conditions to the quantity 
$\langle\xi|\hat W(n,\theta|\xi'\rangle$.  It is sufficient to treat 
$\langle\xi|\hat W(0,\theta|\xi'\rangle$, because of number-shift 
condition (\ref{a22}).  First expand 
$\langle\xi|\hat W(0,\theta|\xi'\rangle)$ as the Fourier series: \zz{b1}
\begin{eqnarray}
 \langle\xi|\hat W(0,\theta)|\xi'\rangle
   &=& \sum_{k,\ell}\langle\xi-\theta|k\rangle
     \langle k|\hat W(0,0)|\ell\rangle\langle\ell|\xi'-\theta\rangle
                                              \nonumber\\
   &=& \frac{1}{2\pi}\sum_k C_k(\xi'-\xi)e^{ik(\theta-\xi)},
                                              \label{b1}
\end{eqnarray}
where \zz{b2}
\begin{equation}
 C_k(\omega)
  = \sum_\ell \langle k+\ell|\hat W(0,0)|\ell\rangle
    e^{i\ell\omega}.                      \label{b2}
\end{equation}
In deriving Eq.\ (\ref{b1}), we have used phase-shift condition 
(\ref{a15}).  It should be noted that the Fourier coefficient 
$C_k(\omega)$ is periodic with a period of $2\pi$.  
Integrating the both sides of equation (\ref{b1}) with respect to 
$\theta$ and using marginal condition (\ref{a13}), we find 
\begin{equation}
  C_0(\omega) = \frac{1}{2\pi}.            \label{b3}
\end{equation}

Since the other marginal condition (\ref{a14}) and number-shift 
condition (\ref{a22}) lead to \zz{b4}
\begin{eqnarray}
 &&\sum_{n=-\infty}^\infty\langle\xi|\hat W(n,\theta)|\xi' \rangle
    = \delta_{2\pi}(\xi-\theta)\delta_{2\pi}(\xi-\xi'), \nonumber\\
 &&\sum_{n=-\infty}^\infty e^{-in(\xi-\xi')}
      \langle\xi|\hat W(0,\theta)|\xi'\rangle
    = 2\pi\delta_{2\pi}(\xi-\xi')
      \langle\xi|\hat W(0,\theta)|\xi'\rangle, \label{b4}
\end{eqnarray}
we arrive at \zz{b5}
\begin{equation}
  \langle\xi|\hat W(n,\theta)|\xi\rangle
    = \frac{\delta_{2\pi}(\theta-\xi)}{2\pi}. \label{b5}
\end{equation}
Similarly, from conditions (\ref{a13}) and (\ref{a17}), it follows 
that \zz{b6}
\begin{equation}
  \langle k|\hat W(n,\theta)|k\rangle = \frac{\delta_{nk}}{2\pi}. 
                                      \label{b6}
\end{equation}
Equation (\ref{b5}) gives \zz{b7}
\begin{equation}
  C_k(0) = \frac{1}{2\pi}.            \label{b7}
\end{equation}

Now let us rewrite overlap condition (\ref{a23}) to obtain another 
condition for the coefficient $C_k(\omega)$.  Noting that \zz{b8}
\begin{eqnarray}
 &&\sum_{n=-\infty}^\infty\int_{-\infty}^\infty\!\!\!du(\theta)\, 
   W(n,\theta)W'(n,\theta)  \nonumber\\
   &&\hspace{5mm}{}= \sum_{n=-\infty}^\infty\int_{-\infty}^\infty\!\!\!
       du(\theta,\xi,\xi',\eta,\eta')\, 
       \langle\xi|\hat W(n,\theta)|\xi'\rangle
       \langle\eta|\hat W(n,\theta)|\eta'\rangle
       \langle\psi|\xi\rangle\langle\xi'|\psi\rangle
       \langle\psi'|\eta\rangle\langle\eta'|\psi'\rangle
                                               \nonumber\\
   &&\hspace{5mm}{}=2\pi\int_{-\infty}^\infty
       \!\!\!du(\theta,\xi,\xi',\eta)\, 
       \langle\xi|\hat W(0,\theta)|\xi'\rangle
       \langle\eta|\hat W(0,\theta)|\xi-\xi'+\eta\rangle \nonumber\\
   &&\hspace{35mm}\times \langle\psi|\xi\rangle
       \langle\xi-\xi'+\eta|\psi'\rangle
       \langle\psi'|\eta\rangle\langle\xi'|\psi\rangle
                                              \label{b8}
\end{eqnarray}
and \zz{b9}
\begin{equation}
 |\langle\psi|\psi'\rangle|^2
  = \int_{-\infty}^\infty\!\!\!du(\xi,\xi',\eta)\, 
    \langle\psi|\xi\rangle\langle\xi-\xi'+\eta|\psi'\rangle
    \langle\psi'|\eta\rangle\langle\eta|\xi'\rangle\langle\xi'|\psi\rangle,
                                              \label{b9}
\end{equation}
we find \zz{b10}
\begin{equation}
 \int_{-\infty}^\infty\!\!\!du(\theta)\, 
 \langle\xi|\hat W(0,\theta)|\xi'\rangle
 \langle\xi'+\zeta|\hat W(0,\theta)|\xi+\zeta\rangle
  = \frac{1}{(2\pi)^2}\langle\zeta|0_{\theta}\rangle, \label{b10}
\end{equation}
where $|0_{\theta}\rangle=|\theta\rangle|_{\theta=0}$.  It follows from 
equation (\ref{b10}) that \zz{b11}
\begin{equation}
 C_k(\omega)C_{-k}(-\omega)=\frac{1}{(2\pi)^2}e^{-ik\omega}.
                                              \label{b11}
\end{equation}

From equation (\ref{b11}) it is convenient to rewrite $C_k(\omega)$ as 
\begin{equation}
 C_k(\omega) =\frac{1}{(2\pi)}e^{-ik\omega/2}\,g_k(\omega).
                                              \label{b12}
\end{equation}
Moreover, equations (\ref{b3}), (\ref{b7}) and (\ref{b12}) 
lead to \zz{b13}
\begin{equation}
  g_0(\omega) = g_k(0) = g_k(\omega)g_{-k}(-\omega) = 1. 
                                              \label{b13}
\end{equation}
It should be noted that we cannot take $g_k(\omega) =1$, because 
$e^{-ik\omega/2}$ does not have a period of $2\pi$ if 
$k$ is odd.  That is, if $k=2m+1$ $(m=0, \pm1, \pm2,\cdots)$, from 
equation (\ref{b12}) it follows that $g_{2m+1}(\omega)$ has period 
$4\pi$: \zz{b14}
\begin{equation}
  g_{2m+1}(\omega +2\pi) = - g_{2m+1}(\omega),\qquad
  g_{2m+1}(\omega +4\pi) = g_{2m+1}(\omega).  \label{b14}
\end{equation}
Then, the quantity $e^{-i\omega/2}\,g_{2m+1}(\omega)$ has a period of $2\pi$ 
and, as a result, the coefficient $C_{2m+1}(\omega)$ has also a period of 
$2\pi$.  

From condition (v), we have \zz{c19}
\begin{equation}
  g_{-k}(-\omega) = g_k(\omega).                  \label{c19}
\end{equation}
Also, it follows from condition (vi) that the function $g_k(\omega)$ 
must be an even function: \zz{c23}
\begin{equation}
  g_k(-\omega) = g_k(\omega).               \label{c23}
\end{equation}

All six conditions give equations (\ref{b13})--(\ref{c23}) for the 
function $g_k(\omega)$.  However, we cannot determine the function 
$g_k(\omega)$ uniquely even if we use these conditions.  Consequently, 
against Ref.\ \cite{biza9455}, the number-phase Wigner operator 
(function) cannot be determined uniquely from the above six conditions.  
If the function $g_k(\omega)$ would not be periodic, then from 
equations (\ref{b13}) and (\ref{c19}), the function $g_k(\omega)$ can 
be determined uniquely; that is, $g_k(\omega)=1$.  In fact, 
Bizarro \cite{biza9455} has been used a solution corresponding 
to $g_k(\omega)=1$.  The origin of the nonuniqueness of definition 
of the Wigner operator is thus the $2\pi-$periodicity of the phase.  

\section{Basic properties of the Wigner operator}

It is shown that the first four conditions, (i) to (iv), are 
sufficient to get basic properties of the Wigner operator; that is, 
these four conditions are fundamental for obtaining the Wigner operator.  
Here, in this section, we do not consider the last two conditions, 
(v) and (vi).  

The number-phase Wigner representation of an operator $\hat A$ is defined 
by \zz{b15}
\begin{equation}
 A(n,\theta)
    \equiv [\hat A](n,\theta)
    = 2\pi\textrm{Tr}[\hat W(n,\theta)\hat A]. \label{b15}
\end{equation}
For the phase operator, from equations (\ref{a7a1}) and (\ref{b5}) and 
the eigenvalue equation $\hat\theta|\xi\rangle=[\xi]|\xi\rangle$, we 
get \zz{b16}
\begin{equation}
 [\hat\theta](n,\theta)
   = 2\pi\int_{-\infty}^\infty\!\!\!du(\xi)\, 
     \langle\xi|\hat W(n,\theta)\hat\theta|\xi\rangle
   = [\theta].                                  \label{b16}
\end{equation}
Similarly, \zz{b17}
\begin{equation}
 [\hat N](n,\theta)
   = 2\pi\sum_{k=-\infty}^\infty
     \langle k|\hat W(n,\theta)\hat N|k\rangle = n, \label{b17}
\end{equation}
where we have used equation (\ref{b6}).  These two Wigner 
representations (\ref{b16}) and (\ref{b17}) are quite natural.  

The trace of a product of any two operators $\hat A$ and $\hat B$ 
can be represented in terms of their Wigner representations: \zz{b18}
\begin{equation}
 \textrm{Tr}[\hat A\hat B]
   = \frac{1}{2\pi}\sum_{n=-\infty}^\infty\int_{-\infty}^\infty
     \!\!\!du(\theta)\, A(n,\theta)B(n,\theta).    \label{b18}
\end{equation}
Indeed, this formula can be derived in a straightforward 
way: \zz{b19}
\begin{eqnarray}
 &&{\hspace{-1cm}}\frac{1}{2\pi}\sum_{n=-\infty}^\infty\int_{-\infty}^\infty
    \!\!\!du(\theta)\, A(n,\theta)B(n,\theta)        \nonumber\\
  &=& 2\pi\sum_{n=-\infty}^\infty\int_{-\infty}^\infty\!\!\!
      du(\theta,\xi,\xi',\eta,\eta')\, 
      \langle\xi|\hat A|\xi'\rangle\langle\eta|\hat B|\eta'\rangle
      \langle\xi'|\hat W(n,\theta)|\xi\rangle
      \langle\eta'|\hat W(n,\theta)|\eta\rangle \nonumber\\
  &=& 2\pi\sum_{k=-\infty}^\infty\int_{-\infty}^\infty\!\!\!
      du(\xi,\xi',\eta)\, 
      \langle\xi|\hat A|\xi'\rangle\langle
      \eta|\hat B|\xi-\xi'+\eta\rangle C_k(\xi-\xi')C_{-k}(\xi'-\xi)
      e^{ik(\xi-2\xi'+\eta)}                    \nonumber\\
  &=& \int_{-\infty}^\infty\!\!\!du(\xi,\xi',\eta)\, 
      \langle\xi|\hat A|\xi'\rangle\langle
      \eta|\hat B|\xi-\xi'+\eta\rangle \delta_{2\pi}(\eta-\xi')
                                                \nonumber\\
  &=& \textrm{Tr}[\hat A\hat B].                 \label{b19}
\end{eqnarray}
Setting $\hat B = |\xi'\rangle\langle\xi|$ in Eq.\ (\ref{b18}), we 
get \zz{b20}
\begin{eqnarray}
 \langle\xi|\hat A|\xi'\rangle
   &=& \sum_{n=-\infty}^\infty\int_{-\infty}^\infty\!\!\!du(\theta)\, 
       A(n,\theta)\textrm{Tr}(\hat W(n,\theta)|\xi'\rangle\langle\xi|)
                                                \nonumber\\
   &=& \sum_{n=-\infty}^\infty\int_{-\infty}^\infty\!\!\!du(\theta)\, 
       \langle\xi|\hat W(n,\theta)|\xi'\rangle A(n,\theta), \label{b20}
\end{eqnarray}
which implies \zz{b21}
\begin{equation}
 \hat A
   = \sum_{n=-\infty}^\infty\int_{-\infty}^\infty\!\!\!du(\theta)\, 
     \hat W(n,\theta)A(n,\theta).               \label{b21}
\end{equation}
From the trace formula (\ref{b18}), it also follows that the expectation 
value of the operator $\hat A$ in a state $\hat \rho$ becomes \zz{b22}
\begin{equation}
  \textrm{Tr}[\hat\rho\hat A]
    = \sum_{n=-\infty}^\infty\int_{-\infty}^\infty\!\!\!du(\theta)\, 
      W(n,\theta)A(n,\theta).                  \label{b22}
\end{equation}

Consider finally the number-phase Wigner function for some simple states.  
For the number state $\hat\rho = |k\rangle\langle k|$, the number-phase 
Wigner function is given by equation (\ref{b6}), whereas the Wigner 
function for the phase state $\hat\rho = |\xi\rangle\langle\xi|$ is 
given by equation (\ref{b5}).

\section{Correct expressions for $\hat S_1(n,\theta)$ and 
$\hat S_2(n,\theta)$}

We present three examples satisfying the first four conditions 
(i) to (iv); the first example corresponds to Vaccaro's operator 
$\hat S_1$ and the second corresponds to $\hat S_2$.  These two examples 
give the correct integral forms for $\hat S_1$ and $\hat S_2$.  To show 
explicitly that the Wigner operator cannot be defined uniquely from 
all six conditions, we consider the third example.  

As the three examples of solutions for $g_k(\omega)$, consider the 
following: \zz{c1,c2,c3}
\begin{eqnarray}
  && \textrm{(a)}\ g_{2m}(\omega) = 1, \quad 
  g_{2m+1}(\omega) \equiv f_1(\omega) = e^{-i\omega/2}; \label{c1}\\
  && \textrm{(b)}\ g_{2m}(\omega) = 1, \quad 
      g_{2m+1}(\omega) \equiv f_2(\omega) = \left\{
                  \begin{array}{rl}
                        1, & \quad(-\pi<\omega<\pi) \\
                       -1, & \quad(\pi<\omega<3\pi);
                  \end{array}
                  \right.                     \label{c2}\\
  && \textrm{(c)}\ g_{4m}(\omega)=1,\quad 
     g_{4m+2}(\omega) = f_2(2\omega),\quad 
     g_{2m+1}(\omega) = f_2(\omega),          \label{c3}
\end{eqnarray}
where $m$ is any integer and $f_2(\omega)$ a square wave with a period 
of $4\pi$.  Note that the first example does not satisfy the last two 
conditions (v) and (vi), whereas the other two ones satisfy all 
conditions (i) to (vi).  

Let us obtain the Wigner operators corresponding to these solutions.  
First consider case (a).  Rewriting $\hat W(n,\theta)$ as 
$\hat W_1(n,\theta)$  and substituting equation (\ref{c1}) into 
equation (\ref{b1}), we find \zz{c4}
\begin{equation}
  \langle\xi|\hat W_1(0,\theta)|\xi'\rangle 
   = \frac{1}{\pi}\delta_{2\pi}(2\theta-\xi-\xi')h_1(\theta-\xi),
                                               \label{c4}
\end{equation}
where \zz{c5}
\begin{equation}
  h_1(\xi) = \frac{1+f_1(2\xi)}{2}.       \label{c5}
\end{equation}
Substituting equations (\ref{a22}) and (\ref{c4}) into equation (\ref{a8}) 
and using equation (\ref{a7a1}), we get \zz{c6}
\begin{eqnarray}
  \hat W_1(n,\theta)
   &=& \frac{1}{\pi}
       \int_{-\infty}^\infty\!\!\!du(\xi,\xi')\, 
       e^{-in(\xi-\xi')}\delta_{2\pi}(2\theta-\xi-\xi')h_1(\theta-\xi)
       |\xi\rangle\langle\xi'|    \nonumber\\
   &=& \frac{1}{\pi}\int_{-\infty}^\infty\!\!\!du(\xi)\, 
       e^{2in(\theta-\xi)}h_1(\theta-\xi)
       |\xi\rangle\langle2\theta-\xi|    \nonumber\\
   &=& \frac{1}{\pi}\int_{-\infty}^\infty\!\!\!d\xi\, U(\theta-\xi)
       e^{2in\xi}h_1(\xi)
       |\theta-\xi\rangle\langle\theta+\xi|. \label{c6}
\end{eqnarray}
Taking equation (\ref{a7a4}) into account, we arrive at the first Wigner 
operator in the extended Fock space: \zz{c8}
\begin{equation}
  \hat W_1(n,\theta)
    = \frac{1}{\pi}\int_{-\infty}^\infty\!\!\!du(\xi)\, e^{2in\xi}
      h_1(\xi)|\theta-\xi\rangle\langle\theta+\xi|.\label{c8}
\end{equation}
By using the projection operator \zz{c9}
\begin{equation}
  P = \sum_{n=0}^\infty|n\rangle\langle n|    \label{c9}
\end{equation}  
onto the physical space spanned by $|n\rangle$ $(n\ge 0)$, the correct 
form of Vaccaro's operator (\ref{a1}) can be derived: \zz{c10}
\begin{equation}
  \hat S_1(n,\theta) = P\hat W_1(n,\theta)P
    = \frac{1}{\pi}\int_{-\infty}^\infty\!\!\!du(\xi)\, e^{2in\xi}
      h_1(\xi)|\theta-\xi;p\rangle\langle\theta+\xi;p|. \label{c10}
\end{equation}
Note that $P|\theta\rangle=|\theta;p\rangle$ [see equation (\ref{a2})].  
Vaccaro's operator (\ref{a1}) should be written as equation 
(\ref{c10}).  Since any physical state $\hat\rho$ satisfies 
$P\hat\rho P = \hat\rho$, we can always use $\hat W_1$ for any 
physical state: \zz{c11}
\begin{equation}
  W_1(n,\theta) 
   \equiv \textrm{Tr}[\hat W_1(n,\theta)\hat\rho]
   = \textrm{Tr}(\hat S_1(n,\theta)\hat\rho)
   \equiv S_1(n,\theta),                       \label{c11}
\end{equation}  
where $S_1(n,\theta)$ is the correct Vaccaro's number-phase Wigner 
function.  

Similarly, in case (b), we have \zz{c12}
\begin{equation}
  \hat W_2(n,\theta)
    = \frac{1}{\pi}\int_{-\infty}^\infty\!\!\!du(\xi)\, e^{2in\xi}
      h_2(\xi)|\theta-\xi\rangle\langle\theta+\xi|,\label{c12}
\end{equation}
where $h_2(\xi)$ is also a square wave, a periodic function with a 
period of $2\pi$: \zz{c13}
\begin{equation}
  h_2(\xi) = \frac{1+f_2(2\xi)}{2}=\left\{
     \begin{array}{rl}
        1, & \quad(-\pi/2<\theta<\pi/2) \\
        0, & \quad(\pi/2<\omega<3\pi/2).
     \end{array}
     \right.\label{c13}
\end{equation}
In deriving equation (\ref{c12}), we have used the equality \zz{c14}
\begin{equation}
  \delta_{2\pi}(2\theta-\xi-\xi')e^{-i(\xi'-\xi)/2}f_2(\xi'-\xi)
   = \delta_{2\pi}(2\theta-\xi-\xi')e^{-i(\theta-\xi)}
     f_2(2(\theta-\xi)).                    \label{c14}
\end{equation}
Note here that $e^{-i\omega/2}f_2(\omega)$ has a period of $2\pi$.  
The physical part of the operator $\hat W_2$ gives the correct expression 
for the second number-phase Wigner operator (\ref{a3}): \zz{c15}
\begin{equation}
  \hat S_2(n,\theta) = P\hat W_2(n,\theta)P
    = \frac{1}{\pi}\int_{-\infty}^\infty\!\!\!du(\xi)\, e^{2in\xi}
      h_2(\xi)|\theta-\xi;p\rangle\langle\theta+\xi;p|. \label{c15}
\end{equation}
In fact, if we consider any ordinary (not generalized) periodic 
function, then equation (\ref{c15}) reduces to equation (\ref{a3}) 
\cite{ligh58,kazu0255}.  That 
is, for any normalizable states $|\psi\rangle$ and $|\phi\rangle$, we 
have \zz{c16}
\begin{eqnarray}
  \langle\psi|\hat S_2(n,\theta)|\phi\rangle 
   &=& \frac{1}{\pi}\int_{-\infty}^\infty\!\!\!du(\xi)\, e^{2in\xi}
      h_2(\xi)\langle\psi|\theta-\xi;p\rangle
      \langle\theta+\xi;p|\phi\rangle.  \nonumber\\
   &=& \frac{1}{\pi}\int_{-\pi/2}^{\pi/2}\!\!\!d\xi\, e^{2in\xi}
      \langle\psi|\theta-\xi;p\rangle
      \langle\theta+\xi;p|\phi\rangle. \label{c16}
\end{eqnarray}
Hence, the operator (\ref{a3}) should be written as 
equation (\ref{c15}).  

The Wigner operator $\hat W_3$ corresponding to case (c) has two 
terms: \zz{c16a}
\begin{eqnarray}
  \hat W_3(n,\theta)
    &=& \frac{1}{\pi}\int_{-\infty}^\infty\!\!\!du(\xi)\, e^{2in\xi}
      h_3(\xi)|\theta-\xi\rangle\langle\theta+\xi|  \nonumber\\
    &&{}+ \frac{1}{\pi}\int_{-\infty}^\infty\!\!\!du(\xi)\, 
      e^{2in\xi}\tilde h_3(\xi)
      |\theta-\pi/2-\xi\rangle\langle\theta-\pi/2+\xi|, \label{c16a}
\end{eqnarray}
where $h_3(\xi)$ and $\tilde h_3(\xi)$ are, respectively, given by \zz{c13}
\begin{equation}
  h_3(\xi) = \frac{1}{2}\left[\frac{1}{2} 
             + \frac{1}{2}f_2(4\xi)+f_2(2\xi)\right],\quad
  \tilde{h}_3(\xi)
   = \frac{1}{4}\left[1 - f_2(4\xi)\right].  \label{c16b}
\end{equation}
In equation (\ref{c16a}) we have used the equality \zz{c16c}
\begin{equation}
  \delta_{2\pi}(2\theta)
   = \frac{1}{2}
     \Big[\delta_{2\pi}(\theta)+\delta_{2\pi}(\theta+\pi)\Big].
                                        \label{c16c}
\end{equation}

The operator $\hat W_1$ does not satisfy conditions (v) 
and (vi), whereas $\hat W_2$ and $\hat W_3$ 
satisfy both of them.  Thus the Wigner operators $\hat W_2$ and 
$\hat W_3$ have higher symmetry than $\hat W_1$.  It should be 
mentioned that there exist infinite Wigner operators satisfying all 
six conditions (i) to (vi).  The operator $\hat W_2$ has the simplest 
integral form.  

To show that the Wigner operators $\hat W_2$ and $\hat W_3$ are better 
than $\hat W_1$, we consider the Wigner representation of the 
operator $\hat\theta\hat N$.  For this purpose, we first get \zz{c24}
\begin{equation}
 \left(\frac{\partial}{\partial\xi}
 \langle\xi'|\hat W(n,\theta)|\xi\rangle\right)_{\xi'=\xi}
 = \frac{i}{2\pi}\left(n-\frac{i}{2}\frac{\partial}{\partial\xi}\right)
   \delta_{2\pi}(\xi-\theta)
   + \frac{1}{(2\pi)^2}\sum_{k=-\infty}^\infty g_k'(0)e^{ik(\theta-\xi)}.
                                              \label{c24}
\end{equation}
Then we arrive at \zz{c25}
\begin{eqnarray}
 [\hat N\hat\theta](n,\theta)
 &=& -2\pi i\int_{-\infty}^\infty\!\!\!du(\xi)\, 
     \left(\frac{\partial}{\partial\xi}
     \langle\xi'|\hat W(n,\theta)|\xi\rangle\right)_{\xi'=\xi}[\xi]
                                              \nonumber\\
 &=& \left(n+\frac{i}{2}\frac{\partial}{\partial\theta}\right)[\theta]
     + \sum_{k=-\infty\atop k\ne0}^\infty\frac{g'_k(0)}{k}e^{ik\theta}
     - \pi ig'_0(0).
                                              \label{c25}
\end{eqnarray}
Since $f'_1(0)=-i/2$ and $f'_2(0)=0$, the number-phase 
Wigner representations for three cases are given by \zz{c26}
\begin{equation}
  \left[\hat N\hat\theta\right](n,\theta)
    = \left\{
    \begin{array}{ll}
     \left(n+\frac{i}{2}{\partial/\partial\theta}\right)[\theta]
        + R(\theta),  & \quad\textrm{case (a)}\\
     \left(n+\frac{i}{2}{\partial/\partial\theta}\right)[\theta],
                      & \quad\textrm{cases (b) and (c)}
    \end{array}
  \right.\label{c26}
\end{equation}
where \zz{c28}
\begin{equation}
  R(\theta) 
    = \sum_{m=0}^\infty\frac{\sin(2m+1)\theta}{2m+1}
    = \frac{\pi}{4}\,f_2(2\theta-\pi/2).      \label{c28}
\end{equation}
The operator $\hat W_1$ leads to a more complex expression for 
$[\hat\theta\hat N](n,\theta)$ than the others.  

Next we show that the operators $\hat W_2$ and $\hat W_3$ correspond to 
Wigner's original operators $\hat W(q,p)$ for position $q$ and momentum 
$p$.  To this end, using $\big[(\hat N\hat\theta)^\dagger\big](n,\theta)
= [\hat N\hat\theta](n,\theta)^*$, we find \zz{c29}
\begin{equation}
  \left[(\hat N\hat\theta+\hat\theta\hat N)/2\right](n,\theta)
    = \left\{
    \begin{array}{ll}
     n[\theta]+R(\theta), & \quad\textrm{case (a)}\\
     n[\theta],           & \quad\textrm{cases (b) and (c).}
    \end{array}
  \right.\label{c29}
\end{equation}
The symmetric operator $(\hat N\hat\theta+\hat\theta\hat N)/2$ has then 
its correspondence $n[\theta]$ in cases (a) and (b).  The operators 
$\hat W_2$ and $\hat W_3$ thus lead to a \lq\lq symmetric" 
representation.  This fact results from $g_k(0)=0$ in the neighborhood of 
the origin.  Recall here that, 
in the original Wigner function, the symmetric operator 
$(\hat q\hat p+\hat p\hat q)/2$ has its correspondence $qp$, where 
$\hat q$ and $\hat p$ are the position and momentum operators, 
respectively.  However, the operator $\hat W_1$ does not have such 
a property.  

\section{Conclusion}

We have investigated the problem of defining the number-phase 
Wigner operator.  We first presented the correct integral expressions 
for the two Wigner operators $\hat S_1(n,\theta)$ and 
$\hat S_2(n,\theta)$, which were derived, respectively, from the 
Wigner operators $\hat W_1(n,\theta)$ and $\hat W_2(n,\theta)$ in 
the extended Fock space.  The operator $\hat W_2(n,\theta)$ satisfies 
all six \lq\lq natural" conditions, whereas $\hat W_1(n,\theta)$ 
satisfies only four ones.  As a result, $\hat W_1(n,\theta)$ does not 
correspond to a symmetric representation; it leads to an unnecessary 
term $R(\theta)$, as shown in equation (\ref{c29}).  The  Wigner 
operator cannot be derived uniquely from the six conditions, because 
of the periodic property of the phase.  To show this fact explicitly, 
we have obtained another Wigner operator $\hat W_3(n,\theta)$ 
satisfying all six conditions, which, howeverr, has more complex 
integral form than $\hat W_2(n,\theta)$.  The operator 
$\hat W_2(n,\theta)$ has the simplest integral form in all other 
Wigner operators.  We need more \lq\lq natural" conditions to define 
the number-phase Wigner operator uniquely, which are not clear at 
present.  

\appendix
\section*{Appendix}

Since it is easy to verify the formulas (\ref{a7a1}) and (\ref{a7a2}), 
we give here proofs of (\ref{a7a3}) and (\ref{a7a4}).  Consider first 
the relation (\ref{a7a3}).  For any states $|\psi\rangle$ and 
$|\varphi\rangle$, we have \zz{x1}
\begin{eqnarray}
  \int_{-\infty}^\infty\!\!\!du(\xi)\, 
  \langle\psi|\xi+\Delta\rangle\langle\xi+\Delta|\varphi\rangle
   &=& \int_{-\infty}^\infty\!\!\!d\xi\, U(\xi-\Delta)
     \langle\psi|\xi\rangle\langle\xi|\varphi\rangle  \nonumber\\
   &=& \sum_{k,\ell=-\infty}^{\infty}
       \langle\psi|k\rangle\langle\ell|\varphi\rangle
       \frac{1}{2\pi}
       \int_{-\infty}^\infty\!\!\!d\xi\, U(\xi-\Delta)e^{i(k-\ell)\xi}
                                            \nonumber\\
   &=& \sum_{k,\ell=-\infty}^\infty 
       \langle\psi|k\rangle\langle\ell|\varphi\rangle
       \frac{1}{2\pi}
       \sum_{m=-\infty}^\infty
       \int_{2\pi m}^{2\pi(m+1)}\!\!\!d\xi\,  
       U(\xi-\Delta)e^{i(k-\ell)\xi}         \nonumber\\
   &=& \sum_{k,\ell=-\infty}^\infty 
       \langle\psi|k\rangle\langle\ell|\varphi\rangle
       \frac{1}{2\pi}
       \sum_{m=-\infty}^\infty
       \int_{0}^{2\pi}\!\!\!d\eta\,  U(\eta-\Delta+2\pi m)
       e^{i(k-\ell)\eta}                       \nonumber\\
   &=& \sum_{k=-\infty}^\infty
       \langle\psi|k\rangle\langle k|\varphi\rangle, \label{x1}
\end{eqnarray}
where we have used $\sum_{m=-\infty}^\infty U(\eta-\Delta+2\pi m) = 1$.  
Equation (\ref{x1}) implies the relation \zz{x2}
\begin{equation}
  \int_{-\infty}^\infty\!\!\!du(\xi)\, 
  |\xi+\Delta\rangle\langle\xi+\Delta|
   = \sum_{k=-\infty}^\infty|k\rangle\langle k| = 1,  \label{x2}
\end{equation}
which is independent of the constant $\Delta$, so that 
equation (\ref{a7a3}) holds.  

Next proceed to prove equation (\ref{a7a4}).  Using the Fourier 
expansion $\langle\xi|\psi\rangle = \sum_{k=-\infty}^\infty 
\langle\xi|k\rangle\langle k|\psi\rangle$, the left-hand side of 
equation (\ref{a7a4}) becomes \zz{x3}
\begin{eqnarray}
  \int_{-\infty}^\infty\!\!\! d\xi\, U(\theta-\xi)\langle\xi|\psi\rangle
   &=& \sum_{k=-\infty}^\infty\int_{-\infty}^\infty\!\!\! d\xi\,  
       U(\theta-\xi)\langle\xi|k\rangle\langle k|\psi\rangle
                                                  \nonumber\\
   &=& \sum_{k=-\infty}^\infty\sum_{\ell=-\infty}^\infty
       \int_{2\pi\ell}^{2\pi(\ell+1)}\!\!\!\!\!d\xi\, U(\theta-\xi)
       \langle\xi|k\rangle\langle k|\psi\rangle  \nonumber\\
   &=& \sum_{k=-\infty}^\infty\sum_{\ell=-\infty}^\infty
       \int_0^{2\pi}\!\!\!d\eta\, U(\theta-\eta-2\pi\ell)
       \langle\eta|k\rangle\langle k|\psi\rangle  \nonumber\\
   &=& \sqrt{2\pi}\langle 0|\psi\rangle.         \label{x3}
\end{eqnarray}
Similarly, it is easily shown that the right-hand side of equation 
(\ref{a7a4}) is also $\sqrt{2\pi}\langle 0|\psi\rangle$.

\end{document}